\documentstyle[11pt,aaspp4,epsfig]{article}

\def\simg{\mathrel{\hbox{\rlap{\lower.55ex \hbox {$\sim$}}
                   \kern-.3em \raise.4ex \hbox{$>$}}}}
\def\siml{\mathrel{\hbox{\rlap{\lower.55ex \hbox {$\sim$}}
                   \kern-.3em \raise.4ex \hbox{$<$}}}}
\def\Mesz{M\'esz\'aros~}
\def\beq{\begin{equation}}
\def\enq{\end{equation}}
\def\bea{\begin{eqnarray}}
\def\ena{\end{eqnarray}}
\def\nonum{\nonumber}
\def\refe{\reference}
\def\bec{\begin{center}}
\def\enc{\end{center}}
\def\etal{{\it et al.}}

\def\cm2si{\hbox{cm$^{-2}$s$^{-1}$}}
\def\cmcui{{\rm cm}^{-3}}
\def\L50{L_{w50}}
\def\E51{E_{w51}}
\def\tw{t_w}
\def\tw1{t_{w1}}
\def\tv3{t_{v-3}}
\def\st{\sigma_T}

\def\eps{\epsilon}

\def\epm{\hbox{e}^\pm}
\def\npm{n_\pm}
\def\Gpm{\Gamma_\pm}
\def\rast{r_\ast}

\begin{document}

\title{ $e^\pm$ Pair Cascades and Precursors in Gamma-Ray Bursts}


\author{P. \Mesz$^{1,2,3}$ , E. Ramirez-Ruiz$^1$, M.J. Rees$^1$}
\smallskip\noindent
$^1${Institute of Astronomy, University of Cambridge, Madingley Road, 
Cambridge
CB3 0HA, U.K.}\\
\smallskip\noindent
$^2${Dpt. of Astronomy \& Astrophysics, Pennsylvania State University,
University Park, PA 16803}\\
$^3${Institute for Theoretical Physics, University of California,
Santa Barbara, CA 93106-4030}

\begin{abstract}
Gamma-ray burst sources with a high luminosity can produce
electron-positron pair cascades in their environment as a result of
back-scattering of a seed fraction of their original hard spectrum.
The resulting spectral modifications  offer the
possibility of diagnosing not only the compactness parameter of
the $\gamma$-ray emitting region but also the baryonic density
of the environment external to the burst.
\end{abstract}


\section{Introduction}

The non-thermal $\gamma$-ray spectrum of Gamma-Ray Burst sources (GRBs)
is thought to arise in shocks which develop beyond the radius at which
the relativistic fireball from the initial event has become optically thin
to scattering.  However, the observed spectra are hard, with a significant
fraction of the energy above the $\gamma\gamma \to \epm$ formation energy
threshold, and a high compactness parameter can result in new pairs being 
formed outside the originally optically thin shocks responsible for the 
primary radiation. New pairs can be made as some of the initial energetic 
photons are backscattered and interact with other incoming photons.
Such effects have been considered by Madau \& Thompson (2000) and Thompson
\& Madau (2000), who investigated the acceleration of new pairs for a 
particular fireball model. Dermer \& B\"ottcher (2000) considered the 
effect of pair formation for an external shock model of GRB,  in which 
the compactness parameter is relatively low, while Madau, Blandford \& 
Rees (2000)  investigated the effects of Compton echos produced by pairs.

Here we present a simplified discussion of the generic effects of pair 
formation in a specific class of models. We suppose that the central
engine gives rise to an unsteady baryonic wind, which is relativistic, 
carries a magnetic field, and lasts for a time $t_w$.  A fraction of the 
wind energy is converted into  $\gamma$-rays  via internal shocks; 
the remaining wind energy drives a blast wave which decelerates as it 
sweeps up the external medium, and gives rise to the afterglow emission.   
The  $\gamma$-rays would propagate ahead of the blast wave, leading to 
pair production (and an associated deposition of momentum)  into the 
external medium.  The pair cascades saturate after the external 
(pair-enriched) medium reaches a critical bulk Lorentz factor, which is 
generally below that of the original relativistic wind. For external 
baryonic densities similar to those in molecular clouds the pairs can 
achieve scattering optical depths $\tau_\pm \siml 1$.  
Even for less extreme external densities the effect of the 
additional pairs can be substantial, increasing the radiative efficiency 
of the blast wave  and leading to distortions 
of the original spectrum. This provides a potential tool  for diagnosing the 
compactness parameter of the bursts and thus the radial distance at which 
shocks producing an observed luminosity can occur. It also provides a tool 
for diagnosing the baryonic density of the external environment of the 
bursts, and testing the association with star-forming regions.

\section{Scattering and Two-photon Pair Formation}

Consider an initial input radiation spectrum produced by the fireball
of the form
\beq
\phi(\eps)= { F \over m_e c^2 \eps_o^2 q } (\eps/\eps_o)^{-\beta}~~
     \cm2si ~~;~~\hbox{for}~~~\eps\geq\eps_o ~,
\label{eq:spin}
\enq
where $F=L/4\pi d_L^2$ is the observed energy flux, $\eps=h\nu/m_e c^2$ is 
photon energy in electron rest-mass units, $\eps_o\sim 0.2-1.0$ is the
break energy above which the photon number spectral index $\beta\sim 2-3$, 
and below which the spectral slope is flatter, e.g. $\alpha\sim -2/3$ for 
a simple low-energy cutoff synchrotron spectrum (although for the present 
purposes the exact low energy slope is unimportant), with $q\sim 1$
a normalization constant. For most values of the low and high energy
slopes, the majority of the photons in the spectrum are near the break 
frequency.

For an $\epm$ moving away radially from the source of radiation with a 
velocity characterized by a Lorentz factor $\Gamma$, taking into account 
the Klein-Nishina drop-off in the scattering cross section above $\eps\sim
1$, the effective fraction of photons contributing to accelerating the
electron is
$\int_{\eps_o}^\Gamma \phi(\eps) d\eps =q^{-1} 
\eps_o^{\beta-2}(\beta-1)^{-1}
[\Gamma^{-1}-\eps_o^{-1}] \simeq q^{-1} \eps_o^{-1} (\beta-1)^{-1}$ for 
$\beta>1$.
At a distance $r$ in front of the radiation source, an $\epm$ can be 
accelerated to a maximum value of $\Gamma$ satisfying
\beq
{L\over 4\pi r^2} {1\over q \eps_o(\beta-1)} (\st /2) \Delta t 
                      \simeq \mu m_e c^2\Gamma~,
\label{eq:scatt}
\enq
where the cross section near $\eps\sim 1$ is approximately $\st/2$,
$\mu m_e$ is the effective mass per scatterer (equation [\ref{eq:mu}]), 
and $\Delta t$ is the effective duration of the light pulse as seen 
by the electron. The latter is either $\Delta t\sim r/2c\Gamma^2$ 
(impulsive regime at small radii), or $\Delta t \simeq t_w$ (wind regime), 
depending on which one is smallest at a given radius, where $t_w$ is the 
duration of the wind, or essentially the burst duration as seen by a 
distant observer. Equation (\ref{eq:scatt}) says that the total 
time-integrated momentum of the radiation intercepted by the electrons 
in the Thompson limit is converted into their kinetic energy of motion, 
taking into account the effective mean mass per scatterer. (This follows 
directly from the mean rate of momentum transfer per particle in a 
steady flow, e.g., as discussed in Madau \& Thompson, 2000, their 
equation (2), (10) and (23), noting that our definition of $\mu$ 
differs from theirs.) 
Here we define $\mu$ as the effective mass per scattering 
electron or positron in units of electron mass,
\beq
\mu m_e ={{\rm mass} \over {\rm scatterer} }=~
 m_p~{ [1+2(\npm/n_p)(m_e/m_p)] \over [1+(2\npm/n_p)]}
 \simeq\cases{ m_p              ~&~ for $\npm\ll n_p$: \cr
               (2\npm / n_p)^{-1}m_p ~&~ for $1<2\npm/n_p<m_p/m_e$;\cr
               m_e              ~&~ for $2\npm/n_p>m_p/m_e$. \cr}
\label{eq:mu}
\enq
Defining a reference radius
\beq
r_\ast= L\st/16\pi m_e c^3=5\times 10^{19}\L50~\hbox{cm}~,
\label{eq:rast}
\enq
from equation (\ref{eq:scatt}) the maximum $\epm$ Lorentz factor is
\beq
\Gpm(r) \simeq \cases{ \mu^{-1/3}
      \left({1 \over q \eps_o (\beta-1) }\right)^{1/3}
      \left({\rast / r }\right)^{1/3} ~,~& for 
                                 $\Delta t\sim ({r / 2c\Gamma^2})$; \cr
      {\mu^{-1}} \left({1\over q \eps_o (\beta-1)}\right)
                           \left({2 c t_w\over\rast}\right)
              \left({\rast / r}\right)^2 ~,~& for $\Delta t\sim t_w$.\cr}
\label{eq:Gammasc}
\enq
The transition from the $\Gamma\propto r^{-1/3}$ to the steeper 
$\Gamma\propto r^{-2}$ occurs at a critical radius $r_{c}\ll \rast$, 
for which the Lorentz factor is $\Gamma_{c}$,
\bea
r_{c}= & 2.2\times 10^{14}\mu^{-2/5}\left(q \eps_o (\beta-1)\right)^{-2/5}
\L50^{2/5}t_w^{3/5}\nonum\\
\Gamma_{c}= & 6.1\times 10^1 \mu^{-1/5}\left(q \eps_o
(\beta-1)\right)^{2/15}
\L50^{1/5}t_w^{-1/5}.
\label{eq:rcsc}
\ena

\begin{figure}[htb]
\centering
\epsfig{figure=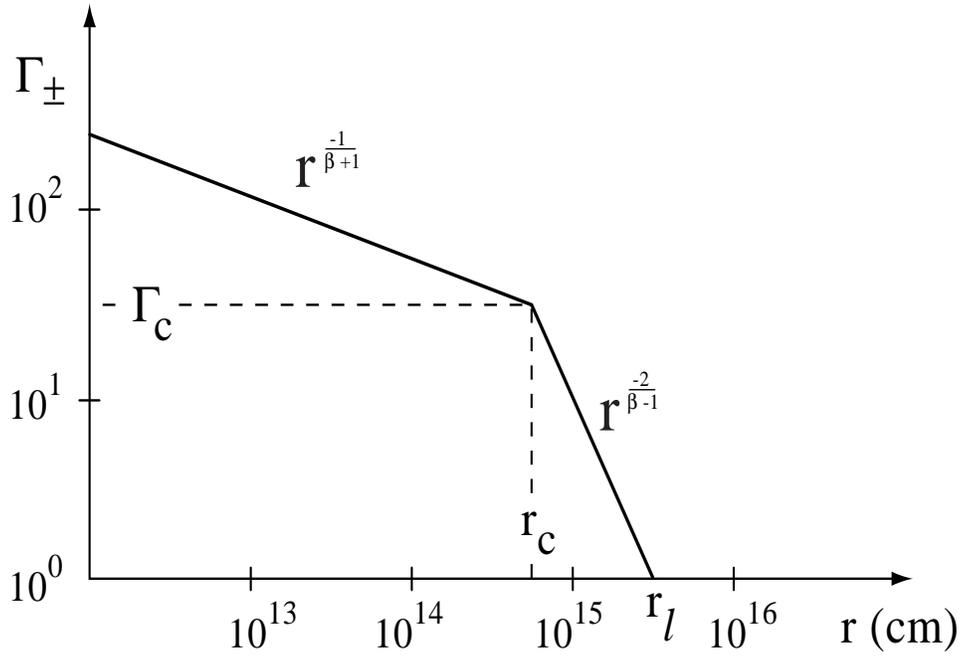,width=5.in,height=3.5in}
\caption{
\small Schematic plot of the maximum $\epm$ bulk Lorentz 
factor $\Gpm$ as a function of radius.
   \label{fig:fig1}
}
\end{figure}

A second criterion for a maximum $\epm$ Lorentz factor comes from the pair 
formation threshold, since the incident photon $\eps$ and the 
back-scattered photon $\eps_r$ must satisfy $\eps\eps_r\geq 2$.
 The back-scattered photon, has,  in the reference frame of the scattering
$\epm$ moving with $\Gpm$, an energy of at most $\eps_r'\siml 1/2$; this
photon cannot give rise to a further pair unless it collides with another
photon with, in the lab frame, an energy exceeding $\sim 4 \Gpm $.
Thus the fraction of incident photons able to make pairs through the 
two-photon mechanism against target photons backscattered  from $\epm$ 
moving with $\Gpm$ is 
$q^{-1}\eps_o^{-2}\int_{4\Gamma}^\infty \phi(\eps)d\eps=
q^{-1}\eps_o^{\beta-2} (\beta-1)^{-1} (4\Gamma)^{1-\beta}$.  
The maximum Lorentz factor achievable by pairs before the two-photon 
cascade cuts off is that for which the compactness parameter has 
dropped to unity,
\beq
\ell (r) = {L\over 4\pi r^2} {\eps_o^{\beta-2} \over q (\beta-1)}
  {\st/3 \over m_e c^2} {1\over (4\Gpm)^{\beta-1}} \Delta t \simeq 1~,
\label{eq:compact}
\enq
where the effective duration is as before.
The maximum Lorentz factor for pair formation is then
\beq
\Gpm(r) \simeq \cases{
     \left({2^{3-2\beta}\over 3(\beta-1)}\right)^{1/{(\beta+1)}}
      \left({\eps_o\over q}\right)^{(\beta-2)/(\beta+1)} 
                                         (\rast / r)^{1/(\beta+1)}
                             ~,~& for $\Delta t\sim ({r / 2c\Gamma^2})$; 
\cr
 \left({4^{2-\beta}\over3(\beta-1)}\right)^{1/(\beta-1)}
  \left({\eps_o\over q}\right)^{(\beta-2)/(\beta-1)}
  \left({c t_w \over\rast}\right)^{1/(\beta-1)}(\rast /r)^{2/(\beta-1)}
                                  ~,~& for $\Delta t\sim t_w$\cr}.
\label{eq:Gammapp}
\enq
This yields a critical radius and Lorentz factor for the transition 
between the impulsive and wind dominated regimes of
\bea
r_{c}= & \left( {2^{7-3\beta} \over 9(\beta-1)^2}\right)^{1\over \beta+3}
  \left({\eps_o\over q}\right)^{2\beta-4\over\beta+3}
  \rast\left({ct_w\over\rast}\right)^{\beta+1\over\beta+3}
 ~\simeq~ 5\times 10^{14}\L50^{2/5}\tw1^{3/5}~,\nonum\\
\Gamma_{c}= & \left({2^{-2(\beta-1)} \over 
3(\beta-1)}\right)^{1\over\beta+3}
  \left({\eps_o\over q}\right)^{\beta-2\over\beta+3}
  \left({\rast\over ct_w}\right)^{1\over\beta+3}
 ~\simeq~ 3\times 10^1 \L50^{1/5}\tw1^{-1/5},
\label{eq:rcpp}
\ena
where the numerical values in the second equation of both lines are 
calculated for $\beta=2$, $\eps_o \simeq q \simeq 1$, $L=10^{50}\L50$ erg 
s$^{-1}$ and $t_w=10\tw1$ s.  (At early times $t<t_w$ the radius $r_c$ 
is smaller, $\propto t^{3/5}$, since it takes $t_w$ for the entire photon 
energy to build up).  The maximum pair Lorentz factor is shown in Figure 
\ref{fig:fig1} for nominal parameter values.  The pair formation limit 
(\ref{eq:rcpp}) is somewhat more restrictive than the scattering limit 
(\ref{eq:rcsc}). Also, unlike the scattering limit, the pair formation 
limit depends exponentially on the photon number slope $\beta$. However,
for the canonical value $\beta=2$ the two power law dependences are the
same and the numerical values are close to each other. In what follows 
we shall use the pair formation limit of equations (\ref{eq:Gammapp}) 
and (\ref{eq:rcpp}).

\section{Pair-precursor and  Kinematics}

Given a certain external baryon density $n_p$ at a radius $r$ outside 
the shocks producing the GRB primordial spectrum, the initial Thomson 
scattering optical depth is $\tau_i \sim n_p\st r$ and a fraction 
$\tau_i$ of the primordial photons will be scattered back, initiating 
a pair cascade. 
Since the photon flux drops as $r^{-2}$, for a uniform (or decreasing) 
external ion density most of the scattering occurs between $r$ and 
$r/2$, and the scattering and pair formation may be approximated as 
a local phenomenon.
This pair-dominated plasma, as long as its density $\npm \siml 
n_p(m_p/2m_e)$, is initially held back by the inertia of its 
constituent ions component, provided that the pairs remain 
coupled to the baryons. (The latter is likely to be the case in
the presence of even weak magnetic fields, e.g. Madau \& Thompson 
2000, or plasma wave scattering, e.g. Lightman, 1982. Moreover, the 
initial magnetic field strength can also increase as consequence of 
instabilities caused by an initial pair streaming relative to ions.
In what follows, we assume that this coupling is effective).

The pairs (together with the ions) start acquiring a significant 
velocity $\Gpm\simg 1$ only after their density $\npm\simg n_p(m_p/2m_e)$,
when the effective mass per scatterer $\mu m_e \sim m_e$.  
After the pairs start being accelerated, 
the pair cascade  can continue multiplying as long as the compactness 
parameter of equation (\ref{eq:Gammapp}) is $\ell\simg 1$.  Both the 
compactness $\ell$ and the maximum $\Gpm$ decrease for increasing 
radius according to equations (\ref{eq:compact}) and (\ref{eq:Gammapp}), 
and the cascade process shuts off for $\ell\siml 1$ at a radius where 
the maximum $\Gpm \sim 1$, 
\beq
r_{\ell}\sim (4 \rast c t_w /3 )^{1/2}\sim 4\times 
10^{15}\L50^{1/2}\tw1^{1/2}~\hbox{cm}.
\label{eq:rell}
\enq
On the other hand, the blast wave producing the shock(s) responsible for 
the afterglow spectrum starts decelerating (i.e. its initial Lorentz 
factor $\Gamma_f\sim\eta =10^2\eta_2$ starts to decrease as a negative 
power law in radius) at the deceleration radius
\beq
r_d\simeq (3 L t_w/4\pi n_p m_p c^2 \eta^2)^{1/3} \simeq
  3\times 10^{15}\L50^{1/3}\tw1^{1/3}\eta_2^{-2/3}n_3^{-1/3}~\hbox{cm}~,
\label{eq:rdec}
\enq
where we normalized to an external baryon density $n_p=10^3n_3\cmcui$. 
(This expression is valid provided that the deceleration radius exceeds 
$ct_w \eta^2$.  Otherwise the deceleration starts before the outflow is 
over, and $t_w$ should be replaced by $t$ in the expression). A larger
value 
$n_p \geq 10^{3}n_3\cmcui$ would give $r_d\siml r_{\ell}$. (Such a higher 
density would, e.g., be required for pair effects to be important in 
an external shock model, where the $\gamma$-rays arise at $r_d$). 
In an internal shock model the primary  $\gamma$-rays themselves arise at 
a smaller radius
\beq
r_i\sim c t_v \eta^2 \sim 3\times 10^{13} t_{v -1} \eta_2^2~\hbox{cm}
\label{eq:rint}
\enq
where $t_v =10^{-1}t_{v-1} \simg 10^{-3}$ s is the variability timescale
in the outflow.  The criterion for an $\epm$ cascade to form ahead of the 
fireball producing a primary spectrum (\ref{eq:spin}) is that $r_i$ (or 
$r_d$ for an external shock model) is smaller than $r_\ell$. 
Pair formation occurs then for $r \leq r_\ell$.

The minimum number of pairs is formed just inside $r_\ell$, where they are 
created essentially at rest, and their density can reach $\npm\sim 
(m_p/2 m_e) n_p \sim 10^3 n_p$, at which point the mass per scatterer has 
become comparable to the electron mass. This represents an increase in the
electron 
(or positron) scattering opacity by a multiplication factor of $k_p\sim 
2\times 10^3$.  As the density exceeds this value at $r_\ell$ the pairs 
are pushed by scattering beyond $r_\ell$, where the condition $\ell <1$ 
prevents further multiplication.

For radii $r<r_\ell$, and as long as the blast wave has not reached that 
radius, pair formation exceeding the above value $\npm\sim 10^3 n_p$ can 
continue to occur, with the pairs being accelerated by scattering until 
they reach the maximum Lorentz factor (\ref{eq:Gammapp}). Pair production 
will be most copious at smaller radii $r_i <r < r_{c}$, where the pairs 
are accelerated to a maximum Lorentz factor $\Gpm(r)\siml 
\Gamma_{c}(r/r_{c})^{-1/3}$.  At each step of the cascade the Lorentz 
factor of the new generation of pairs approximately doubles, and the 
maximum pair multiplication factor of the cascade can be estimated from 
the number $s$ of pair generations required to go from $\Gamma\sim 1$
to $\Gpm(r)$ given by equation (\ref{eq:Gammapp}). At $r_{c}$ this is
$s(r_c) \sim \log \Gamma_c /\log 2 \sim 5$, and for $r_i \siml r $ it is 
limited to $s(r_i) \siml 7.5 $. This represents an extra multiplication 
factor $k_a \sim 2^s \sim \Gamma_c (r/r_c)^{-1/3} \sim 30-170$ in the 
electron opacity for $r_i \leq r \leq r_c$, in addition to the previous
factor 
$k_p \sim 2\times 10^3$ achieved before acceleration starts. The critical
value is
\beq
k_c = k_p k_a(r_c) \sim (m_p/m_e) \Gamma_c \sim 
  5\times 10^4 \L50^{1/5}\tw1^{-1/5}~,
\label{eq:kc}
\enq
with $k(r)=k_c (r/r_c)^{-1/3}$ for $r_i\leq r_c$, while $k(r) \sim k_c
(r/r_c)^{-2}$ 
for $r_c\leq r \leq r_\ell$ drops to $m_p/m_e$ at $r_\ell$, and to
zero beyond that. In a more detailed calculation, the number of
cascades depends somewhat on the radiation spectral index $\beta$,
here set equal to 2. For a steeper spectrum, the number of cascades is 
reduced ( e.g. Ramirez-Ruiz {\it et al.}  2000). 
The maximum pair optical depth is achieved at $r_c$,
\beq
\tau_{\pm c} \sim k_c n_p \st r_c =
 \min [~2\times 10^{-5} n_p \L50^{3/5}\tw1^{2/5},~1~].
\label{eq:tauc}
\enq
The pair opacity scales as $\tau_\pm\propto r^{2/3}$ for $r_i\siml r \siml
r_c$
and $\tau_\pm\propto r^{-1}$ for $r_c\siml r \siml r_\ell$.
For an external density greater than
\beq 
n_{p,c}\simeq 10^5 \L50^{-3/5} \tw1^{-2/5} \cmcui ~, 
\label{eq:npc}
\enq
the pair density inside $r_c$ is prevented from multiplying beyond a value 
corresponding to $\tau_\pm\sim 1$, due to self-shielding.
The number of pairs created at $r_c$ is
\beq
N_{\pm c}\sim n_{p} k_c (4\pi/3)r_c^3 \sim 3\times 10^{49} n_p \L50^{7/5}
\tw1^{8/5}~,
\label{eq:Npmc}
\enq
which scales as $N_\pm\propto r^{8/3},~r$ for $r$ below or above $r_c$.
The maximum value for $n_p=n_{p,c}$ is 
$N_{\pm,c,mx}\sim 3\times 10^{54}\L50^{4/5} \tw1^{6/5}$ at $r=r_c$, and 
$N_{\pm,\ell ,mx}\sim 3\times 10^{55}\L50^{9/10} \tw1^{11/10}$ at
$r_\ell$.
The pair cloud extends over a range of radii comparable to $r$, much
larger 
than the photon pulse radial width $c t_w$. 

\begin{figure}[htb]
\centering
\epsfig{figure=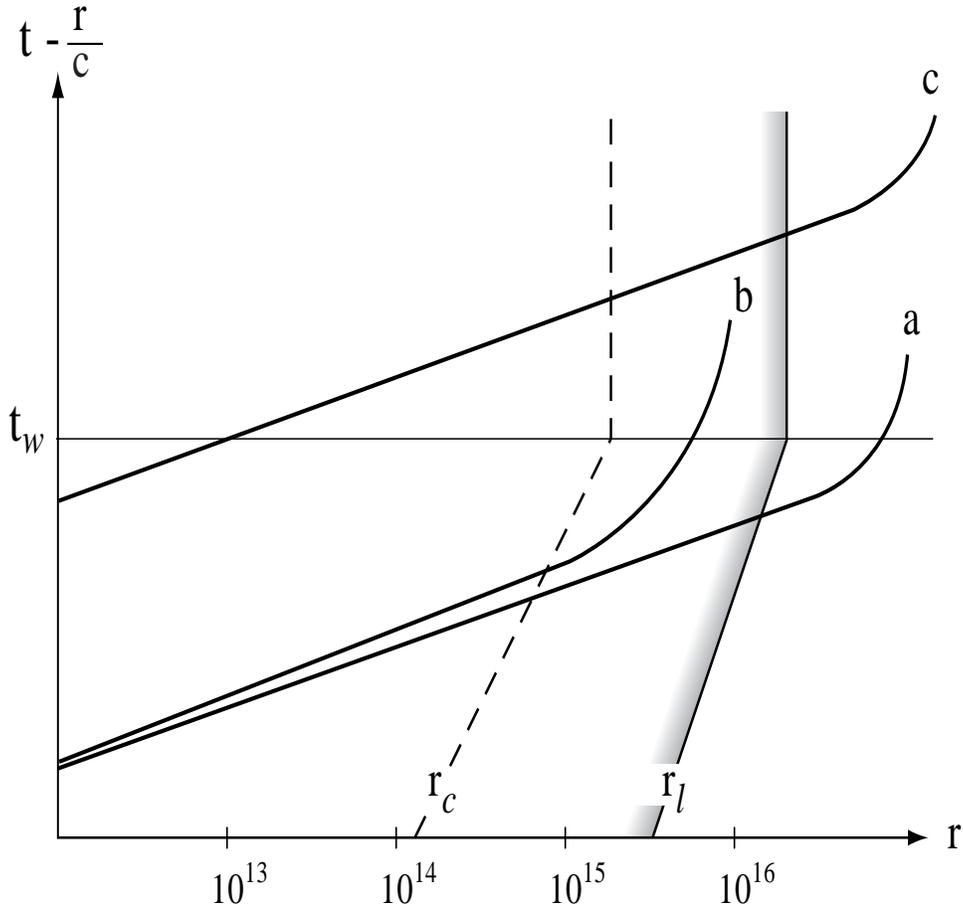,width=5.in,height=5.in}
\caption{
\small
This diagram shows, for three illustrative cases, how the burst would be
affected by pair-enrichment. The axes (logarithmic) are {\it r} versus $t-
(r/c)$, where $t$ is time measured by a distant observer, and is zero when
the burst is observed to start. (In this plot, light rays are horizontal
lines). The primary gamma-ray emission is assumed to continue, with a
quasi-steady luminosity {\it L}, for a time $t_w$ (as would happen if, for
instance, it came from internal shocks induced by wind variability on
time-scales $\ll t_w$). The criterion for runaway pair production would be
satisfied within the ambient material out to a radius $r_\ell$. (cf
equation
[10];  {\it t} replaces $t_w$ in this equation for times $t < t_w$.) The
associated absorption of momentum would itself accelerate this
pair-enriched material to an $r$-dependent Lorentz factor (see Figure
1). The wind generally carries more momentum than the gamma rays, and
drives a
relativistic blast wave that eventually sweeps up all the pair-enriched
medium.
    Three illustrative cases are depicted. In case (a), the external
medium has a low density, and the blast wave, with high $\eta$, sweeps up
all the pair-enriched medium before it has been much decelerated: the 
effects of the pairs are then observed primarily during the burst itself.
   In case (b), with higher external density, deceleration occurs at radii
$< r_\ell$, and the blast wave is still moving through pair-enriched
material
during the afterglow. When the ambient medium is dense, the pairs may
provide an optical depth of unity, so the primary burst itself would be
reprocessed, and its short time-structure smeared out.
  Case (c) corresponds to a lower $\eta$. The sweeping-up of pairs then
occurs during the afterglow (modifying the radiative efficiency of the
outer shock) even if the external density is low and there has not (as in
case b)  already been deceleration.
   \label{fig:fig2}
}
\end{figure}

Figure \ref{fig:fig2}  shows the schematic world-lines of three blast
waves
expanding into an external medium.  At each radius $r_i < r < r_\ell$
pairs
form ahead of $r_i$.  While for $r\sim r_\ell$ the 
pairs are sub-relativistic and do not propagate significantly, for smaller 
radii they move relativistically with an initial speed $\Gpm$, with 
$\Gpm<\eta$ in general, although at very small radii, and for times 
$t<t_w$,  $\Gpm> \eta$ may be possible.
If the blast wave did not exist, then the
pair-enriched material (moving out with a Lorentz factor that is larger at
smaller r) would pile up at a radius $r_l$ after a time $r_l/c$. However,
the pair-rich external medium would generally carry  less energy and 
inertia than the relativistic wind itself; it therefore  starts being 
decelerated by the external medium at a
smaller radius than a blast wave with the same Lorentz factor would be, so
that the  latter always overtakes it and sweeps  up the pair-enriched 
external medium. 
The resultant outer shock is more likely to be radiative, but may be
weaker, because the ambient material might already be moving outward
relativistically before the blast wave hits it.

For a relatively low external density such as $n_p=10^3\cmcui$, as in
curve  (a) of Figure \ref{fig:fig2}, the radius $r_c <r_d$, and the 
pairs are swept up by a fireball expanding at its original  speed 
$\eta\sim$ constant, and deceleration of the fireball begins sometime
after the pairs have been swept up. In this case the main afterglow, 
produced by the deceleration of the pair-enriched fireball will be 
characterized by a larger emission measure than in the usual case, 
due to the extra pairs in the swept-up matter, but will not otherwise 
exhibit any irregularities in its time development.
For a larger external density such as $10^6\cmcui$, corresponding to
curve (b) of Figure \ref{fig:fig2}, the radius $r_c\sim r_d$ and the
main part of the afterglow, caused by the deceleration of the fireball 
baryons and the pairs formed at early stages, continues to sweep up 
further pairs which were formed at or beyond the deceleration radius 
(in curve (c), with low $\eta$, deceleration occurs at radii 
$r>\;r_\ell$). 
These new pairs change the emission measure of the deceleration shock, 
since their optical depth $\tau_\pm\sim 1$ is larger than that of the 
swept up baryonic electrons.
If $\tau_\pm \sim 1$ is reached, the original spectrum may be modified by
Comptonization, and may approach a quasi-thermal shape. We note that the
optical depth cannot substantially exceed unity. This is because
high-energy  non-thermal photons  would be comptonised by pairs , and
after a few scatterings would be degraded below the pair-production
threshold. So, however intense the
radiation incident on its inner edge,  a shell of  pair-dominated plasma
cannot build up an opacity much larger than unity. It will come to a
quasi-steady state in which the luminosity from its outer surface
(comptonised synchrotron radiation and annihilation radiation) equals the
non-thermal radiation shining on its inner surface.

\section{Observable Pair-precursor Effects}
\label{sec:prec}

The energy in the accelerated pairs is $E_\pm \propto N_\pm \Gamma_\pm r^3
\propto 
r^{7/3}$ for $r\siml r_c$ and $E_\pm \propto r^{-1}$ for $r_c\siml r \siml
r_\ell$.
The maximum number of pairs created is
\beq
E_{\pm c} \simeq  N_\pm m_e c^2\Gamma_c = 
\min[~10^{45} n_p \L50\tw1 ~,~ 10^{50} \L50^{2/5}\tw1^{3/5}]~{\rm erg},
\label{eq:Epm}
\enq
where the second number is the limit $N_{\pm ,mx}$ obtained for 
$n_p=n_{p c}$, at which the saturation value of $\tau_\pm\sim 1$ is 
obtained. The pair energy never exceeds more than a fraction of the 
initial wind energy, so the deceleration radius (equation [\ref{eq:rdec}]) 
is unaffected, while the pairs which are produced and accelerated  ahead 
of the blast wave, are decelerated at a (smaller) radius
\beq
r_{d\pm}\sim 2.5\times 10^{15} E_{\pm 50}^{1/3} 
n_3^{-1/3}(\Gpm/30)^{-2/3}~{\rm cm}~,
\label{eq:rdecpm}
\enq
before the ejecta itself starts to decelerate.

The total number of protons in the ejecta wind is  $N_p =E/\eta m_p c^2= 
(2/3)\times 10^{52}\L50\tw1\eta_2^{-1}$. The number of protons in the 
(pair enriched) external medium which are swept up by the time the blast 
wave reaches its deceleration radius $r_d$ is $N_{ps}=N_p\eta^{-1}= 
(2/3)\times 10^{50}\L50\tw1 \eta_2^{-2}$. The maximum number of swept-up 
pairs can exceed the number of baryons by a large factor. However the 
inertia of the pairs is less than that of the baryonic ejecta, since
$(N_{\pm,c,mx}/N_{p})\sim 5\times 10^2 \L50^{-1/5}\tw1^{1/5}\eta_2 \siml 
(m_p/m_e)(\eta/\Gamma_c) =6\times 10^3 \L50^{-1/5}\tw1^{1/5}\eta_2$ at
$r_c$, and $(N_{\pm,\ell, mx}/N_p)\sim 5\times 10^3\L50^{-1/10}
\tw1^{1/10}\eta_2 \leq (m_p/m_e)\eta=2\times 10^5\eta_2$ at $r_\ell$.

The pairs will modify the usual properties of the  deceleration shocks and
the afterglow emission. The total energy available in the afterglow is not
changed: it is still, essentially, the kinetic energy of the relativistic 
wind, minus the fraction dissipated (and converted into prompt gamma rays)  
in internal shocks. But the lepton/proton ratio in the ejecta can be  much 
larger than usual. This increases the radiative efficiency significantly, 
since most of the particles  are $\epm$.   The  resultant radiation will
thus be softer than in the usual picture, because the same energy density 
has to be shared among a larger number of particles ($N_\pm/N_p \gg 1$).
In the reverse shock that occurs after the baryonic+pair ejecta starts
being decelerated by the external medium, the comoving frame peak random
electron Lorentz factor is 
$\gamma_{\pm, m,r} \sim (E \eta^{-1}/ N_\pm m_e c^2)$.
Even for $\tau_\pm$ substantially less than unity, e.g. for an external 
density $n=10^3 n_3\cmcui$ (below the critical value $n_{p,c}$) for which
the deceleration radius $r_d\sim r_c$, the reverse shock random lepton 
Lorentz factor is $\gamma_{\pm, m, r} \sim 
30\E51^{-1/5}\tw1^{-3/5}n_3^{-1} \eta_2^{-1}$. The pairs should not
affect, however, the random turbulent magnetic field strength, which is 
in pressure equilibrium with the forward shock at some fraction $\eps_B$ 
of the equipartition value, $B'\sim (\eps_B 8\pi n m_p c^2)^{1/2}\eta 
\sim 6\times \eps_B^{1/2} n_3^{1/2}\eta_2$ G.
Thus the observed reverse synchrotron peak frequency 
\beq
\nu_{\pm,sy,r}\sim 10^6 B'\gamma^2 \eta (1+z)^{-1} \sim 6\times 10^{13}
\E51^{-2/5}\tw1^{-6/5}\eps_B^{1/2} n_3^{-3/2}\eta_2^{-1} (1+z)^{-1}~{\rm
Hz}
\label{eq:nusyr}
\enq
would be in the far IR, as opposed to the optical/UV of the usual 
baryon-dominated prompt reverse flash.

The forward shock afterglow radiation, normally in the $\gamma$/X-ray 
range, would be unaffected outside $r_\ell$. However, up to about the
time when deceleration  starts, there will be a ``pair pick-up" 
photon pulse, when the protons and electrons of the relativistic wind 
moving with $\eta >\Gamma_c$  sweep up the pair-enriched external 
medium. The protons are not decelerated by the pairs, so there is only 
a subsonic reverse compression wave in the proton ejecta,  but there 
will be a mildly relativistic forward shock in the picked-up pairs.
The $\epm$ random Lorentz factor will be  equal to the bulk kinetic
energy it  has in the proton frame, e.g. $\gamma_{\pm, m, f} \sim
\eta/\Gpm \sim 3 \L50^{-1/5}\tw1^{1/5}\eta_2$ near $r_c$. The comoving
random magnetic field  in the shocked pair fluid is $B'\sim 
(8\pi \eps_B N_{\pm,c} m_e c^2 /4 r_c^3)^{1/2} (\eta/\Gamma_\pm)$ 
$\sim 20 \eps_B^{1/2}\L50^{1/5}\tw1^{1/10} n_3^{1/2}\eta_2$ G for 
$n_p=10^3 n_3\cmcui$, corresponding to a pair-pickup pulse synchrotron 
peak frequency in the observer frame of
\beq
\nu_{\pm,sy,f} \sim 2\times 10^{10} 
  \eps_B^{1/2}\L50^{-1/5}\tw1^{5/10}n_3^{1/2}\eta_2^3 (1+z)^{-1}~{\rm Hz},
\label{eq:nusyf}
\enq
and a synchrotron power law above this frequency.

There could also be additional pair-precursor signatures which are not 
associated with the ejecta blast wave, but with the dynamics of the 
pre-accelerated pair-enriched plasma.  The r-dependent  $\Gpm$ of the 
pair-enriched external plasma will lead to internal shocks.  Pair regions 
at $r<r_c$ whose $\Gpm$ differ by order unity will collide at radii 
$\propto r \Gpm^2 \propto r^{1/3}$, so for sufficiently short variability 
times these shocks would occur between $r_i$ and $r_c$, i.e. up to an 
observer time $t_c\sim r_c/c\Gamma_c^2\sim 15 \tw1$ s, independent of the 
wind luminosity $L$.  (For $r_c <r <r_\ell$ the collision radii are 
$\propto r^{-3}$, so the shocks would tend to pile up at $r_\ell$ at an 
observer timescale $t_\ell\sim r_\ell/c \sim$ 1 day, provided the main
part of the fireball wind has not caught up with it before. This could 
be the case for a jet-like wind for pairs in a rim of angular width
$\Gamma_c^{-1}$ around the jet).  
The pair internal shocks at $r\sim r_c$ can produce a Fermi accelerated 
power law $\epm$ spectrum above $\gamma_{\pm m} \sim 1$ and a turbulent 
comoving field $B'_{\pm } \siml 2\times 10^2$ G, leading to a 
synchrotron spectrum whose observer-frame peak frequency $\nu_{\pm m} 
\sim 6\times 10^9$ Hz would be self-absorbed, with a power law extension 
above it which would be optically thin at higher frequencies. 

The above observational signatures would be present even if $\tau_\pm$
is low, as expected for external ion densities $n_p \siml n_{p,c} \sim 
10^5 \L50^{-3/5} \tw1^{-2/5} \cmcui$ around the burst. An additional
effect of interest, for external densities in excess of this which 
lead to a pair screen of optical depth $\tau_\pm \sim 1$, is that a 
quasi-thermal  pulse of X-rays could accompany the burst, caused by 
upscattering of diffuse progenitor stellar photons.  
A collapsing massive progenitor leading to a GRB is likely to be highly 
super-Eddington for sometime after the GRB event, e.g. 
$L_\ast \sim 10^4 L_{Ed}\sim 10^{43} L_{\ast 43}$ erg/s. 
The density of photons of energy $\eps_\ast\sim 10$ eV near $r_c$, which 
would be quasi-isotropized due to the condition $\tau_\pm \sim 1$ at 
$r_c$ (and beyond, where $\Gamma_\pm \to 1$ at $r_\ell$), would 
be $n_\ast \sim 10^{13} L_{\ast 43} \eps_{\ast 10}^{-1}\L50^{-4/5} 
\tw1^{-6/5}\cmcui$ near $r_c$. The screen with $\tau_\pm\sim 1$ moving 
with $\Gamma_c\sim 30$ sweeps up a total number of photons $N_{\ast}\sim 
3\times 10^{57}L_{\ast 43}\eps_{\ast 10}^{-1} \L50^{2/5}\tw1^{3/5}$. The 
mean energy per photon and the time-integrated total energy of the
upscattered precursor is 
\bea
\eps_X\sim & 10~\eps_{\ast 10} \L50^{2/5}\tw1^{-2/5}(1+z)^{-1}
      ~{\rm keV}\nonum\\
E_X\sim & 5\times 10^{49} L_{\ast 43}\eps_{\ast 10}^{-1}\L50^{4/5}
  \tw1^{1/5} (1+z)^{-1}~{\rm erg}.
\label{eq:xrprec}
\ena
This would last until the pairs are swept up by the ejecta,
$t_X \siml t_c\sim r_c/c\Gamma_c^2\sim 15 \tw1$ s.

\section{Discussion}

We have discussed, in the context of a standard internal/external shock 
model of gamma-ray bursts (which are normally assumed to occur after the
original fireball has become optically thin) the  $\gamma\gamma\to \epm$ 
cascades triggered by the back-scattering of seed gamma-ray photons on 
the external medium. This effect can modify the initial scattering optical 
depth of the outflow at radii $r\siml r_\ell \simeq 4\times 
10^{15}\L50^{1/2}\tw1^{1/2}~\hbox{cm}$ (equation [\ref{eq:rell}]), which
is comparable to the radii of external shocks of equation (\ref{eq:rdec}) 
at which the afterglow begins, and is generally larger than the typical 
internal shock radii given by equation (\ref{eq:rint}).  The spectral 
effects of the pairs  on the burst and the afterglow can be substantial, 
and within radii $\sim r_l$ they can affect the dynamics.

Pair production can increase  the optical depth outside of the shocks 
by  up to  $\siml 10^5$;  self-shielding ensures that the maximum
scattering optical depths achievable by the pairs is $\tau_\pm\sim 1$. For 
typical interstellar densities the pair opacity $\tau_\pm \ll 1$, which
does not significantly affect the gamma-ray spectrum. The number of pairs 
may nonetheless be large enough to increase the radiative efficiency and 
soften significantly the radiation spectrum of the afterglow reverse
shock, where the same energy is shared among a number of $\epm$ which 
can exceed that of the original $e^-$ and $p^+$ of the ejecta.

The  pair production processes themselves (determining $r_c,~ r_l$, 
equations [\ref{eq:rcpp}], [\ref{eq:rell}]) just depend on the ``seed" 
$\gamma$-ray photon flux (which are here postulated to come from internal 
shocks).  The manifestations depend on the external density and on the 
initial dimensionless entropy or bulk Lorentz factor $\eta$. The external 
baryon density $n_{ext}$ determines the optical depth that can be
built up through back-scattering and pair multiplication. This affects 
whether the pair optical depth gets up to unity, with smearing and 
reprocessing of the primordial $\gamma$-ray spectrum, or whether it merely 
makes the blast wave more radiative. Madau and Thompson (2000) have made 
this point, in the context of a specific fireball model.  

The external density (along with the initial Lorentz factor $\eta$) 
determines when the  outer shock and the reverse shock become important 
and whether this happens within the radius already polluted with
pairs  (and pre-accelerated by radiation pressure before the shock hits).
There are two rather different cases depending on whether or not $\eta^2$ 
is less than $r_l/ct_w$). In the former case the external shock 
responsible for the afterglow occurs beyond the region ``polluted" by 
new pairs, while in the second case the afterglow shock may experience, 
after starting out in the canonical manner, a ``resurgence" or second 
kick as its radiative efficiency is boosted by running into an 
$\epm$-enriched gas.  Internal shocks in the pair-dominated external
plasma  can lead to self-absorbed radiation at $\sim 10^9-10^{10}$ Hz, 
while the swept-up pairs can also contribute  a $10^{11}-10^{12}$ Hz 
`prompt' signal, which precedes the onset of the standard deceleration 
afterglow phase.

Additional effects are expected when $\tau_\pm \to 1$. This requires 
external baryon densities at radii $r<r_\ell$ of $n_p \simg n_{c,p} \sim  
10^5 \L50^{-2/5} \tw1^{-3/5} \cmcui$. Such high densities would 
only be expected if the burst is associated with a massive star in which 
prior mass loss  led to  a dense circumstellar envelope  The pair optical 
depth saturates to $\tau_\pm\sim 1$ and in addition to an increased 
efficiency and softer spectrum of the afterglow reverse shock, the
original gamma-ray spectrum of the GRB will be modified as well.  The 
specific nature of this spectral modification depends on the value of 
the luminosity, which influences (equation [\ref{eq:rcpp}]) the bulk 
Lorentz factor of the reprocessing pair cloud before it has been swept 
up by the ejecta, and also on the extent to which the outflow is beamed. 
One of the consequences of such a critical external density leading to
$\tau_\pm\sim 1$ would be the presence of an X-ray quasi-thermal pulse, 
whose total energy may be a few percent of the total burst energy. 
In the case of even more extreme densities, there are other interesting 
possibilities. For instance, dense blobs of Fe-enriched thermal plasma 
would emit strong recombination features, as well as annihilation 
radiation features, if the normal electron density were augmented by 
extra pairs. 

       Even in isotropic situations, the spectrum would be modified  by 
transmission through a pair plasma of optical depth unity. The effect is 
maximal for photons of energy $m_e c^2$ in the frame of the pair plasma:
for higher energies, the Klein-Nishina cross-section is smaller; for 
lower energies, the scatterings are almost elastic.  If, for instance, 
$\Gamma_c\sim 40$, the gamma-ray spectrum around photon energies 
$\eps\sim \Gamma_c m_e c^2 \sim 20$ MeV  would be depressed by a factor
$\sim 1$ relative to its original value, smoothly rejoining its original 
value above and below that energy.
  For a beamed primary output, however, there would be a suppression at 
lower energies (where the scattering is in the Thompson regime) because
the scattered photons would be spread over a wider angle.
Pair-induced processes would therefore yield evidence on the beaming 
properties of the bursts.

Irrespective of the external density, the processes discussed here suggest 
that bursts and afterglows may have a more complex spectrum and 
time-structure than 'standard' models suggest. But the effects are 
especially interesting when the external density is high: they probe 
the environment of GRBs, and thus can offer clues to the nature of the 
progenitor  stars,  and their location within the host galaxy.
For instance, a quasi-thermal X-ray pulse accompanying the gamma-ray 
emission could be indicative of an external circum-burst density of at 
least $10^4-10^5~\cmcui$. While quasi-thermal X-ray pulses might also 
arise due to other reasons, e.g. from an underlying optically thick 
central engine, if the X-ray luminosity scales as equation 
(\ref{eq:xrprec}) and is accompanied by radio or far-IR signals 
such as in equations (\ref{eq:nusyr})(\ref{eq:nusyf}), this could be
indicative of birth in a dense environment from a massive progenitor.

Detailed Monte Carlo simulations (Ramirez-Ruiz \etal, 2001) should provide
a more detailed assessment of the self-consistent spectrum of a GRB in
the presence of self-induced pair formation.

\acknowledgements{This research has been supported by NASA NAG5-9192, the 
Guggenheim Foundation, the Sackler Foundation, NSF PHY94-07194 and the 
Royal Society. We are grateful to R. D. Blandford, C. Dermer, P. Madau, 
C. Thompson and the referee for useful  comments.}


\end{document}